\documentclass[onecolumn,preprintnumbers,prd,superscriptaddress,nofootinbib]{revtex4}
\usepackage{amsmath}
\usepackage{amsfonts}
\usepackage{amssymb}
\usepackage{graphicx}
\usepackage{color}
\usepackage{comment}

\usepackage{amsmath,amssymb,amsthm}
\usepackage{physics}
\usepackage{bm}
\usepackage{graphicx}
\usepackage{hyperref}
\def\be{\begin{equation}}
\def\ee{\end{equation}}
\def\bea{\begin{eqnarray}}
\def\eea{\end{eqnarray}}

\begin{document}

\title{\textbf{Thermodynamic acceptability of spherically symmetric perfect-fluid solutions in general relativity}}

\author{Dina~\surname{Demissenova}}
\email[]{dina.d.a.94@inbox.ru}
\affiliation{Al-Farabi Kazakh National University, Al-Farabi av. 71, 050040 Almaty, Kazakhstan.}

\author{Nurzada~\surname{Beissen}}
\email[]{nurzada.beissen@gmail.com}
\affiliation{Al-Farabi Kazakh National University, Al-Farabi av. 71, 050040 Almaty, Kazakhstan.}

\author{Kuantay~\surname{Boshkayev}}
\email[]{kuantay@mail.ru}
\affiliation{Al-Farabi Kazakh National University, Al-Farabi av. 71, 050040 Almaty, Kazakhstan.}

\author{Hernando {Quevedo}}
\email[]{quevedo@nucleares.unam.mx}
\affiliation{Instituto de Ciencias Nucleares, Universidad Nacional Aut\'onoma de M\'exico, 04510 Mexico City, Mexico }
\affiliation{Dipartimento di Fisica and ICRA, Universit\`a di Roma “La Sapienza”, Roma, Italy}
\affiliation{Al-Farabi Kazakh National University, Al-Farabi Ave., 71, Almaty, 050040, Kazakhstan}

\begin{abstract}

Static spherically symmetric perfect-fluid solutions of Einstein's equations play a central role in relativistic astrophysics and stellar structure theory. While many exact solutions satisfy Einstein's equations mathematically, only a limited subset satisfies physically acceptable conditions such as regularity, positivity of matter variables, and causal sound propagation. In this work, the classical concept of physical acceptability is extended to include thermodynamic considerations. Using relativistic equilibrium thermodynamics, entropy functionals, and the Tolman temperature relation, we formulate a set of thermodynamic acceptability conditions for relativistic stellar models. The Tolman IV solution is analyzed as an explicit example. We show that this solution admits a finite and positive equilibrium entropy functional consistent with the Tolman equilibrium condition. This analysis suggests that thermodynamic consistency provides a natural extension of the Delgaty-Lake acceptability program and may constitute an essential criterion in the classification of relativistic interior solutions.

\end{abstract}

\maketitle
\section{Introduction}

Static, spherically symmetric perfect-fluid solutions of Einstein’s gravitational field equations have long occupied a position of fundamental importance in relativistic astrophysics, gravitation theory, and the study of compact astronomical objects. These solutions provide some of the most important theoretical models for understanding how matter behaves under the influence of extremely strong gravitational fields, particularly in regimes where Newtonian gravity is no longer adequate and the full framework of general relativity becomes necessary.

Since the formulation of general relativity by Einstein in 1915, considerable effort has been devoted to the search for exact interior solutions describing self-gravitating fluid distributions in equilibrium. Such solutions are intended to represent the internal structure of compact astrophysical systems, including relativistic stars, radiation-dominated spheres, ultra-dense stellar remnants, and other highly compressed matter configurations. Because these objects possess approximately spherical symmetry and are often modeled as isotropic fluids in hydrostatic equilibrium, static spherically symmetric perfect-fluid models provide a natural and mathematically tractable framework for their investigation.

Over the decades, a large number of exact interior solutions have been constructed in an attempt to describe different physical properties of relativistic matter and stellar structure. These models have played a central role in exploring the consequences of Einstein’s equations in strong-field regimes and in developing theoretical insight into the behavior of dense gravitational systems. Classical examples include the Schwarzschild interior solution \cite{Tolman1939}, the Tolman family of solutions \cite{Tolman1939}, Buchdahl models \cite{Buchdahl1959,Buchdahl1967}, Adler solutions \cite{Adler1974}, and Heintzmann configurations \cite{Heintzmann1969,HeintzmannHillebrandt1975} among others \cite{Oppenheimer1939,FinchSkea1989,Bondi1964,MakHarko2003,Ivanov2002}. 

A major issue in the study of exact solutions is that mathematical consistency alone does not guarantee physical relevance. Einstein's equations admit a large number of exact solutions that may exhibit unphysical features such as singular densities, negative pressures, superluminal sound speeds, or the absence of a finite stellar surface. Consequently, the study of exact solutions in general relativity gradually evolved beyond the mere construction of mathematically consistent models. Increasing attention began to be devoted to determining whether these solutions could genuinely represent physically realistic self-gravitating systems. As a result, researchers placed growing emphasis on analyzing the physical plausibility of exact solutions, including their regularity, stability, causal behavior, and compatibility with the known properties of relativistic matter.

An especially influential contribution to the study of realistic relativistic stellar models was made by Delgaty and Lake \cite{Delgaty1998}, who carried out a systematic investigation of the physical acceptability of static, spherically symmetric perfect-fluid solutions of Einstein’s equations. Their work represented an important development in relativistic astrophysics because it emphasized that the mathematical existence of an exact solution is not sufficient to guarantee its physical relevance. Instead, they argued that any solution intended to model a realistic compact object must satisfy a number of additional physical requirements associated with regularity, causality, and hydrostatic stability.

In their analysis, Delgaty and Lake formulated a practical framework for determining whether a given exact interior solution could reasonably describe a self-gravitating relativistic star. They emphasized that physically acceptable configurations should possess regular behavior at the stellar center, so that quantities such as the metric functions, pressure, and density remain finite and free from singularities. Furthermore, both the pressure and energy density must remain positive throughout the interior region of the star, since negative matter variables would generally lack physical interpretation in ordinary stellar matter.

They also required that the density and pressure decrease monotonically outward from the center toward the surface, reflecting the physically expected behavior of stable stellar configurations in hydrostatic equilibrium. Another essential requirement was the existence of a finite boundary radius at which the pressure vanishes, thereby allowing the interior fluid distribution to match smoothly onto an exterior vacuum spacetime. In addition, the sound speed inside the fluid must remain subluminal in order to preserve relativistic causality, implying that pressure disturbances cannot propagate faster than the speed of light. Finally, the solution must satisfy the conditions of hydrostatic equilibrium as determined by the Einstein equations and the relativistic stellar structure equations.

Their investigation demonstrated that only a relatively small subset of the many known exact perfect-fluid solutions actually satisfy all of these elementary physical requirements simultaneously. This result had important consequences for the study of exact solutions in general relativity because it established the notion of physical acceptability as a fundamental criterion in relativistic stellar modeling. From that point onward, the evaluation of exact solutions increasingly focused not only on mathematical solvability, but also on their consistency with physically realistic properties of self-gravitating matter.
However, the Delgaty--Lake program was concerned primarily with geometric regularity and hydrodynamic viability, and did not systematically incorporate thermodynamic considerations into the analysis of exact interior solutions. Although their criteria successfully identified many solutions that were mathematically and physically reasonable from the standpoint of relativistic stellar structure, important thermodynamic questions remained largely unexplored.

In particular, comparatively little attention was devoted to determining whether a given exact solution admits a well-defined and physically meaningful entropy functional, or whether the associated temperature distribution is compatible with the conditions required for relativistic thermal equilibrium in curved spacetime. Similarly, the extent to which exact perfect-fluid solutions satisfy the laws of relativistic thermodynamics was not investigated in detail. Another important unresolved issue concerns the microscopic interpretation of the matter source itself, namely whether the fluid described by a given exact solution can be related in a consistent way to realistic microphysical models of dense matter, radiation, or nuclear interactions occurring in compact astrophysical objects.

The principal objective of the present work is to extend the traditional notion of physical acceptability in relativistic stellar structure so as to incorporate thermodynamic considerations in a systematic manner. More specifically, we investigate whether exact perfect-fluid solutions that satisfy the usual geometric and hydrodynamic requirements of relativistic astrophysics may also be regarded as thermodynamically acceptable configurations within the framework of general relativity. The analysis is motivated by the observation that a stellar model may exhibit mathematically regular behavior and satisfy standard physical conditions, while nevertheless failing to possess a consistent thermodynamic interpretation.

Our approach is based on several fundamental ingredients of relativistic thermodynamics, including the Gibbs relation, entropy currents, and the Tolman equilibrium condition governing temperature distributions in curved spacetime. Particular attention is devoted to the distinction between local thermodynamic consistency and genuine global thermal equilibrium, since these notions are not equivalent in the presence of gravitational fields. In relativistic systems, local equations of state and locally defined temperatures do not automatically guarantee equilibrium throughout the entire spacetime, because gravitational redshift modifies the thermal structure of self-gravitating matter distributions.

Within this framework, we formulate a set of thermodynamic acceptability conditions intended to complement the classical Delgaty--Lake criteria. These conditions include the positivity and regularity of entropy density, the finiteness of total entropy, compatibility with the relativistic Gibbs relation, positivity of temperature, causal propagation, and consistency with the Tolman equilibrium law. The purpose of these requirements is to identify exact interior solutions that are not only geometrically and hydrodynamically viable, but also thermodynamically meaningful.

As an explicit example, we investigate the Tolman IV solution, one of the best-known exact perfect-fluid solutions of Einstein’s equations. This solution is already known to satisfy many of the standard conditions of physical acceptability and therefore provides a natural framework in which to explore the thermodynamic aspects of relativistic stellar models. By applying the Tolman temperature relation and constructing the associated entropy functional, we show that the Tolman IV configuration admits a finite, positive, and regular equilibrium entropy throughout the stellar interior. The analysis further demonstrates that the corresponding temperature profile is compatible with relativistic thermal equilibrium and that the entropy remains well behaved under physically reasonable parameter choices.

The results obtained in this work suggest that thermodynamic consistency provides a natural and physically important extension of the classical acceptability program for exact solutions in general relativity. More generally, the analysis indicates that realistic relativistic stellar models should be evaluated not only in terms of geometric regularity and hydrodynamic stability, but also with respect to their compatibility with equilibrium thermodynamics and microscopic physical interpretation.

This work is organized as follows. In Sec. \ref{sec:pf}, we review the main aspects of spherically symmetric perfect-fluid spacetimes in general relativity. Then, in Sec. \ref{sec:rtd}, we include some aspects of relativistic thermodynamics and present the conditions that must be satisfied for a perfect-fluid interior solution to be meaningful from a thermodynamic perspective. Section \ref{sec:tolmaniv} is devoted to the study of thermodynamic acceptability of the Tolman IV spacetime. Finally, in Sec. \ref{sec:con}, we  discuss our results and comment on some open questions.  


\section{Static perfect-fluid configurations}
\label{sec:pf}

We consider static, spherically symmetric spacetime geometries describing the interior gravitational field of a relativistic fluid distribution in equilibrium. Owing to the assumptions of time independence and spherical symmetry, the spacetime metric may be written in Schwarzschild-like coordinates, $x^a=(t,r,\theta,\phi)$ in the general form

\be
ds^2
=-e^{\nu(r)}dt^2
+
e^{\lambda(r)}dr^2
+
r^2 (d\theta^2+\sin^2\theta d\phi^2) ,
\label{metric}
\ee 
where the metric functions $\nu(r)$ and $\lambda(r)$ depend only on the radial coordinate $r$.  The assumption of staticity implies the existence of a timelike Killing vector field, while spherical symmetry guaranties that the geometry is invariant under spatial rotations. These symmetries considerably simplify the Einstein field equations and make static perfect-fluid configurations one of the most extensively studied classes of exact solutions in general relativity.

The matter source is assumed to be a perfect fluid characterized by isotropic pressure and energy density. The corresponding stress-energy tensor is therefore taken to be
\be
T_{ab}
=
(\rho+p)u_a u_b
+
p g_{ab},
\ee 
where $\rho(r)$ denotes the energy density measured by comoving observers, $p(r)$ is the isotropic pressure, and $u^a$ is the four-velocity of the fluid elements. In the static case, the fluid remains at rest with respect to the chosen coordinate system, so that the four-velocity satisfies
\be
u^a = e^{-\nu/2}\delta^a_0,
\label{4vel}
\ee 
together with the normalization condition
\be
u^a u_a = -1.
\label{norm}
\ee 

Under these assumptions, Einstein’s field equations reduce to a coupled system of ordinary differential equations relating the metric functions $\nu(r)$ and $\lambda(r)$ to the thermodynamic variables $\rho(r)$ and $p(r)$. The resulting equations can be written as 
\be
\rho(r)= \frac{1}{8\pi r^2 e^\lambda}\left(r\lambda' + e^\lambda-1\right),
\label{eins1}
\ee
\be
p(r) = \frac{1}{8\pi r^2 e^\lambda}\left( 
r\nu' - e^\lambda +1\right),
\label{eins2}
\ee
\be
2\nu'' + (\nu')^2 - \nu'\lambda' - \frac{2}{r}(\lambda'+\nu') +  \frac{4}{r^2}(e^\lambda -1) = 0,
\label{eins3}
\ee
determine the internal structure of the relativistic fluid sphere and encode the condition of hydrostatic equilibrium through the Tolman--Oppenheimer--Volkoff equation \cite{Tolman1939,Oppenheimer1939}.

A physically realistic stellar configuration must possess a finite boundary radius $r=R$ at which the pressure vanishes, $p(R)=0.$
At this surface, the Darmois-Israel matching conditions \cite{Israel1966} should be satisfied, which imply that the first fundamental form 
\be
h_{ij} = g_{ab} \frac{\partial x^a}{\partial \xi^i}\frac{\partial x^b}{\partial \xi^j}
\ee
and the second fundamental form 
\be
K_{ij} = \nabla_a n_b \frac{\partial x^a}{\partial \xi^i}\frac{\partial x^b}{\partial \xi^j}
\ee
should be continuous along the matching surface $\Sigma: r=R$, i.e., 
\be
[h_{ij}] = h^+_{ij} - h_{ij}^- =0,\quad
[K_{ij}] = K^+_{ij} - K_{ij}^- =0.
\ee
Here, $\xi^i=(t,\theta,\phi)$ are the coordinates of the matching surface, $\nabla_a$ denotes the covariant derivative, $n_a = e^{\lambda(R)/2}\delta_a^r$ represent the components of the unit normal vector to $\Sigma$, and the index $\pm$ denotes the corresponding quantity outside ($+$) and inside ($-$) the matching surface $\Sigma$. For the general line element (\ref{metric}), we obtain 
\be
h_{tt} = - e^{\nu(R)}, \quad 
h_{\theta\theta}= R^2, \quad
h_{\phi\phi}= R^2 \sin^2\theta
\ee
\begin{equation}
K_{tt}=-\frac{1}{2}e^{\nu(R)-\lambda(R)/2}\nu'(R),\quad
K_{\theta\theta}=Re^{-\lambda(R)/2},\quad
K_{\phi\phi}=Re^{-\lambda(R)/2}\sin^2\theta .
\end{equation}
Then, from the firs matching condition $[h_{ij}]=0$, we obtain that 
\be
e^{\nu_-(R)} = e^{\nu_+(R)}.
\label{di1}
\ee
Furthermore, from the second matching condition $[K_{ij}]=0$, taking into account the continuity of $\nu(R)$ as given in (\ref{di1}), we obtain
\be
e^{\lambda_-(R)}=e^{\lambda_+(R)},\quad
\nu'_-(R)=\nu'_+(R).
\label{di2}
\ee
Moreover, imposing the matching with the exterior vacuum Schwarzschild solution, for which 
\be
e^{\nu_+(r)} = 1 - \frac{2M}{r}, \quad 
e^{-\lambda_+(r)}= 1 - \frac{2M}{r},
\ee
 Using the matching condition for the function $e^{\lambda(r)}$
as given in Eq. (\ref{di2}), together with the Einstein equation (\ref{eins2}), one can see that condition $\nu'_-(R)=\nu'_+(R)$ is equivalent to condition $p(R)=0$. Consequently, we have shown that using the Einstein field equations, the continuity of the extrinsic curvature is equivalent to the vanishing of the pressure at the stellar surface, $p(R)=0$. Therefore, for the Schwarzschild exterior spacetime, the Darmois–Israel matching conditions reduce to the continuity of the metric functions $e^\nu$ and $e^\lambda$, together with the condition $p(R)=0$ \cite{Israel1966,Stephani2003,Quevedo2021}.

In this work, as mentioned above, we will impose the physical condition $p(R)=$ to define the matching radius. Consequently, to satisfy the the Darmois-Israel matching conditions we only need to demand the continuity of the metric functions, i.e.,
\be
e^{\nu(R)} =
1-\frac{2M}{R}, \ \ 
e^{-\lambda(R)}
=
1-\frac{2M}{R}.
\label{match}
\ee 

These junction conditions ensure the absence of surface layers or singular shells at the boundary and guarantee that the interior fluid distribution joins continuously onto the external vacuum geometry. The complete relativistic stellar model is therefore determined by solving Einstein’s equations inside the fluid region together with the appropriate boundary and matching conditions at the stellar surface.


\section{Relativistic thermodynamics}
\label{sec:rtd}

In relativistic thermodynamics, one assumes that matter is in local thermodynamic equilibrium. This means that although the thermodynamic variables may vary from one spacetime point to another because of gravity and spatial inhomogeneity, each infinitesimal fluid element may still be described by the ordinary laws of equilibrium thermodynamics. Consequently, quantities such as temperature, entropy density, pressure, chemical potential, and particle number density can be defined locally in the instantaneous rest frame of the fluid.

The starting point is the first law of thermodynamics applied to a relativistic fluid element. Let the total internal energy be denoted by $U$, the entropy by $S$, the volume by $V$, and the total particle number by $N$. The ordinary  first law of thermodynamic is
\be
dU = T\, dS - p\, dV + \mu\, dN,
\ee 
where $T$ is the temperature, $p$ is the pressure, and $\mu$ is the chemical potential associated with particle conservation. 
To obtain a local relativistic form, we introduce the corresponding densities measured in the local rest frame of the fluid: $U=\rho V,
\
S=sV,
\
N=nV.$ 
Using  these definitions, from the first law we obtain the Euler identity 
\be
\rho + p = Ts + \mu n,
\ee 
and the relativistic Gibbs relation 
\be
d\rho = Tds + \mu dn.
\ee 
This relation expresses local energy conservation for a relativistic fluid element and constitutes the fundamental thermodynamic identity used throughout relativistic fluid theory. The Euler identity relates the entropy density directly to the macroscopic fluid variables appearing in Einstein’s equations. In relativistic thermodynamics, the combination $\rho+p$ plays a central role not only in hydrodynamics and gravitational equilibrium, but also in the thermodynamic description of matter.

In the special case where the chemical potential vanishes, $\mu = 0$, the Euler identity simplifies considerably. This situation commonly occurs for radiation fluids and photon gases because photons are not conserved particles and therefore possess zero chemical potential in equilibrium \cite{Weinberg2008}. The resulting identity $s = \frac{\rho+p}{T}$
provides a direct connection between thermodynamic entropy and the gravitational fluid variables that characterize the stellar interior. 
To construct the total entropy of a static relativistic star, one must integrate the local entropy density over the proper spatial volume of the configuration. For a static spherically symmetric configuration (\ref{metric}), we obtain
\be
    S = \int s dV_{proper} = 
4\pi
\int_0^R
s(r)e^{\lambda/2} r^2 dr = 
4\pi
\int_0^R
\frac{\rho+p}{T(r)}
e^{\lambda/2} r^2 dr.
\label{entropy}
\ee 
This expression defines the natural entropy functional for static relativistic perfect-fluid configurations. It incorporates both the local thermodynamic properties of the matter and the geometric effects of spacetime curvature through the metric coefficient $e^{\lambda/2}$. Consequently, the total entropy depends simultaneously on the thermodynamic state of the fluid and on the gravitational structure determined by Einstein’s equations.

An important point should be emphasized at this stage: the calculation of the entropy functional requires explicit knowledge of the temperature profile throughout the stellar interior. Unlike the pressure and energy density, which are determined directly from Einstein’s equations once a particular exact solution is specified, the temperature is not automatically fixed unless additional thermodynamic conditions are imposed.

Consequently, the entropy cannot be computed solely from the geometric variables or the hydrodynamic quantities of the solution. We must first determine a physically consistent temperature distribution $T(r)$. In relativistic systems, this issue is particularly important because thermal equilibrium in curved spacetime is governed by the Tolman relation rather than by a spatially constant temperature.

Thermal equilibrium in general relativity differs fundamentally from its Newtonian counterpart because gravitational fields redshift energy. As a consequence, a system in equilibrium cannot generally have a uniform local temperature throughout the gravitational field. Instead, the equilibrium condition derived by Tolman requires that the redshifted temperature remain constant across the spacetime \cite{Tolman1930}. The Tolman equilibrium condition is
$ T(r)\sqrt{-g_{tt}(r)}
= const. $ for a spherically symmetric spacetime (\ref{metric}) reads
\be
T(r)e^{{\nu(r)}/{2}} = T_\infty,
\label{temp}
\ee 
where $T_\infty$ is a constant redshifted temperature measured by an observer at infinity.

The physical meaning of this condition is important. It guaranties the absence of heat flow within the configuration and ensures that no net thermal energy is transported between different fluid layers. In addition, the Tolman relation implies vanishing entropy production in equilibrium and therefore characterizes a state of global thermal equilibrium in curved spacetime. For this reason, the Tolman condition may be regarded as the relativistic generalization of the zeroth law of thermodynamics \cite{Tolman1930,TolmanEhrenfest1930,Tolman1934,Eckart1940,Israel1979}.

Configurations that fail to satisfy this condition generally correspond to non-equilibrium systems \cite{LandauLifshitz1987}. In such cases, temperature gradients are not properly compensated by gravitational redshift effects, and the system may exhibit heat transport, dissipative processes, and nonvanishing entropy production.

An important distinction must therefore be made between local thermodynamic consistency and true global equilibrium. A local equation of state may remain perfectly meaningful even when the global equilibrium condition is violated. For example, the ideal-gas relation
\be
p = nk_B T,
\ee 
may still hold locally within each infinitesimal fluid element. In this situation, standard thermodynamic identities such as the Gibbs relation continue to be valid locally, and quantities such as pressure, density, temperature, and entropy density may all remain well defined and positive throughout the stellar interior. Nevertheless, the overall configuration does not represent an exact equilibrium state unless the Tolman relation is satisfied simultaneously.

Consequently, entropy constructed from a locally defined temperature profile that violates the Tolman condition should be interpreted as a non-equilibrium entropy functional rather than an equilibrium entropy. By contrast, entropy obtained using the Tolman temperature possesses a fundamentally different interpretation because it corresponds to a genuine equilibrium configuration in curved spacetime.

Once the Tolman temperature profile has been specified, the entropy functional may be evaluated unambiguously using (\ref{entropy}).
Thus, the entropy depends not only on the matter variables $\rho$ and $p$, but also on the gravitationally determined temperature profile required by relativistic thermal equilibrium. The Tolman condition therefore plays a central role in connecting thermodynamics with the geometry of self-gravitating relativistic systems.

\subsection{Thermodynamic acceptability conditions}

Motivated by the program initiated by Delgaty and Lake concerning the physical acceptability of exact relativistic stellar models, we propose that physically realistic perfect-fluid solutions should also satisfy a corresponding set of thermodynamic acceptability conditions. The purpose of these requirements is to ensure not only that the spacetime geometry and hydrodynamic variables behave in a physically reasonable manner, but also that the underlying thermodynamic description remains internally consistent and compatible with relativistic equilibrium thermodynamics.

The first requirement is that the entropy density remain positive throughout the stellar interior,
\be
s(r)>0.
\ee 
This condition reflects the fundamental thermodynamic interpretation of entropy as a measure of microscopic disorder or the number of accessible microscopic states. Negative entropy densities would lack a meaningful physical interpretation and would signal a breakdown of the thermodynamic description of the matter source.

A second requirement is that the total entropy of the configuration be finite,
\be
0<S<\infty.
\ee 
A finite entropy ensures that the stellar configuration possesses a well-defined global thermodynamic state. Divergent entropy would generally indicate pathological behavior such as singular thermodynamic variables, infinite proper volume contributions, or the breakdown of equilibrium thermodynamics near certain regions of spacetime.

The temperature profile must also remain positive everywhere inside the fluid sphere,
\be
T(r)>0.
\ee 
Negative temperatures are not appropriate for ordinary relativistic stellar matter and would correspond to highly exotic nonequilibrium systems outside the scope of the present analysis. The positivity of temperature is therefore regarded as a basic requirement for thermodynamic viability.

In addition, the temperature distribution should satisfy the Tolman equilibrium condition, as  explicitly given in Eq.(\ref{temp}). This condition guarantees that the system is in genuine thermal equilibrium in curved spacetime. In relativistic gravitating systems, local temperature gradients are unavoidable because of gravitational redshift effects, and equilibrium is achieved not through spatially constant temperature, but through the constancy of the redshifted temperature. Satisfaction of the Tolman relation therefore ensures the absence of heat flow and vanishing entropy production.

However, it should be emphasized that other temperature profiles may also be introduced depending on the assumed microscopic interpretation of the fluid. For example, we may define a local temperature through the ideal-gas relation $p=nk_B T$,
or through more complicated microphysical equations of state. Such temperature definitions may remain locally meaningful and thermodynamically consistent within each infinitesimal fluid element. Nevertheless, a locally defined temperature obtained in this way does not necessarily satisfy the Tolman equilibrium condition globally. Consequently, the resulting configuration may correspond to a non-equilibrium relativistic fluid with heat transport and entropy production. In the present work, the Tolman temperature is adopted because the primary goal is to investigate equilibrium thermodynamic acceptability in curved spacetime.

Another important requirement is compatibility with the relativistic Gibbs relation,
\be
Tds=d\rho-\mu dn.
\ee 
This condition guarantees that the thermodynamic variables admit a consistent local thermodynamic interpretation. The Gibbs relation provides the fundamental differential identity connecting entropy density, energy density, temperature, chemical potential, and particle number density. A physically meaningful stellar model should therefore allow its matter variables to satisfy this relation throughout the interior region.

The entropy density and all associated thermodynamic quantities must furthermore remain regular and free from singularities inside the stellar configuration. In particular, no divergences should occur at the stellar center or at any finite interior radius. The absence of entropy singularities is closely related to the regularity conditions usually imposed on the metric functions, pressure, and energy density in physically acceptable exact solutions.

In addition to thermodynamic consistency, the fluid must also satisfy the standard relativistic causality condition on the sound speed,
\be
0<\frac{dp}{d\rho}<1.
\ee 
This inequality ensures that acoustic perturbations propagate at speeds smaller than the speed of light. Violation of this condition would imply either mechanical instability or superluminal propagation, both of which are physically unacceptable within relativistic fluid theory.
It is worth noting that this requirement is not an additional independent assumption in our framework but is already included within the original Delgaty–Lake physical acceptability criteria. In their analysis of static, spherically symmetric perfect-fluid solutions, this condition is explicitly imposed as part of the standard set of requirements for a solution to be considered physically realistic. It therefore belongs to the hydrodynamic and stability constraints originally used to classify acceptable relativistic stellar models, and in the present work it is simply retained and complemented by the additional thermodynamic conditions introduced here.

Finally, the thermodynamic variables should exhibit monotonic behavior throughout the stellar interior. In physically realistic compact objects, one generally expects quantities such as temperature, pressure, density, and entropy density to decrease smoothly outward from the center toward the stellar surface. Nonmonotonic behavior may indicate instabilities, phase transitions, or the presence of unphysical matter distributions.

Taken together, these conditions provide a thermodynamic extension of the classical geometric and hydrodynamic acceptability criteria introduced by Delgaty and Lake. They are intended to identify exact relativistic solutions that are not only mathematically regular and gravitationally stable, but also compatible with the fundamental principles of equilibrium thermodynamics in curved spacetime.


\section{The Tolman IV Solution}
\label{sec:tolmaniv}

As an explicit application of the thermodynamic acceptability criteria introduced in the previous sections, we now consider the Tolman IV solution. This solution was originally obtained by Tolman in his classical investigation of static perfect-fluid spheres in general relativity and remains one of the most widely studied exact interior solutions of Einstein’s equations. Its importance arises from the fact that, for suitable choices of the parameters, it satisfies many of the standard conditions of physical acceptability required for realistic relativistic stellar models. In particular, the solution possesses regular metric functions, positive pressure and density, a finite boundary radius, and causal sound propagation.

\subsection{Metric and fluid variables}

The spacetime geometry of the Tolman IV solution is described by the line element (\ref{metric})
with the metric functions given by
\be
e^{\nu}
=
B^2\left(1+r^2/A^2\right),
\ee 
and
\be
e^{-\lambda}
=
\frac{
\left(1-r^2/C^2\right)
\left(1+r^2/A^2\right)
}{
1+2r^2/A^2
},
\ee 
which satisfy Einstein equations (\ref{eins1})-(\ref{eins3}) with
the energy density 
\be
\rho(r)
=\frac{3A^4+2r^2(3r^2+C^2)+A^2(7r^2+3C^2)}{8\pi C^2(A^2+2r^2)^2},
\ee 
and the isotropic pressure 
\be
p(r)
=
\frac{C^2-A^2-3r^2}{8\pi C^2(A^2+2r^2)}.
\ee 
The constants $A$, $B$, and $C$ characterize the geometry of the stellar interior and determine the behavior of the gravitational potentials. 
These quantities remain finite at the stellar center. In particular, the central density and pressure are
\be
\rho(0)
=\frac{3(A^2+C^2)}{8\pi A^2 C^2},
\ee 
and
\be
p(0)
=
\frac{C^2-A^2}{8\pi A^2C^2}.
\ee 
Thus, positivity of the central pressure requires
\be
C^2>A^2.
\ee 

The stellar surface is defined by the radius $r=R$ at which the pressure vanishes, $
p(R)=0.$ This condition implies
\be
C^2=A^2+3R^2,
\label{pzero}
\ee 
alternatively
\be
R=\sqrt{(C^2-A^2)/3}.
\ee 

Moreover, the mass profile turns out to be 
\be
m(r)=4\pi\int_0^r \tilde{r}^2\rho(\tilde{r}) d\tilde{r} = \frac{r^3
   \left(A^2+C^2+r^2\right)}{2 C^2
   \left(A^2+2 r^2\right)}.
\ee 

Consequently, the configuration possesses a finite radius and may be smoothly matched to the exterior Schwarzschild vacuum solution
at the boundary surface $r=R$. The matching conditions require continuity of the metric coefficients, which using the zero-pressure condition (\ref{pzero}) leads to 
\be 
M=m(R)=\frac{R^3}{C^2}=\frac{R^3}{A^2+3R^2}, 
\quad
B^2 = \frac{A^2}{A^2+3R^2} = 1-\frac{3M}{R}.
\ee
From these equations one can express constants $A$ and $C$ in terms of the total mass and radius of a compact object:
\be 
C^2=\frac{R^3}{M}, 
\quad
A^2 = \frac{R^2(R-3M)}{M} .
\ee
The matter variables exhibit physically reasonable behavior throughout the stellar interior. Both the pressure and energy density decrease monotonically away from the center, while the metric functions remain regular and free from singularities for $0\leq r\leq R$. Furthermore, for suitable parameter values, the sound speed
\be
v_s^2
=
\frac{dp}{d\rho},
\ee 
remains smaller than the speed of light everywhere inside the star. Consequently, the Tolman IV solution satisfies the classical physical acceptability conditions proposed in the Delgaty-Lake proposal. 
.

\subsection{Entropy functional }

\begin{figure}
    \centering
    \includegraphics[scale=0.4]{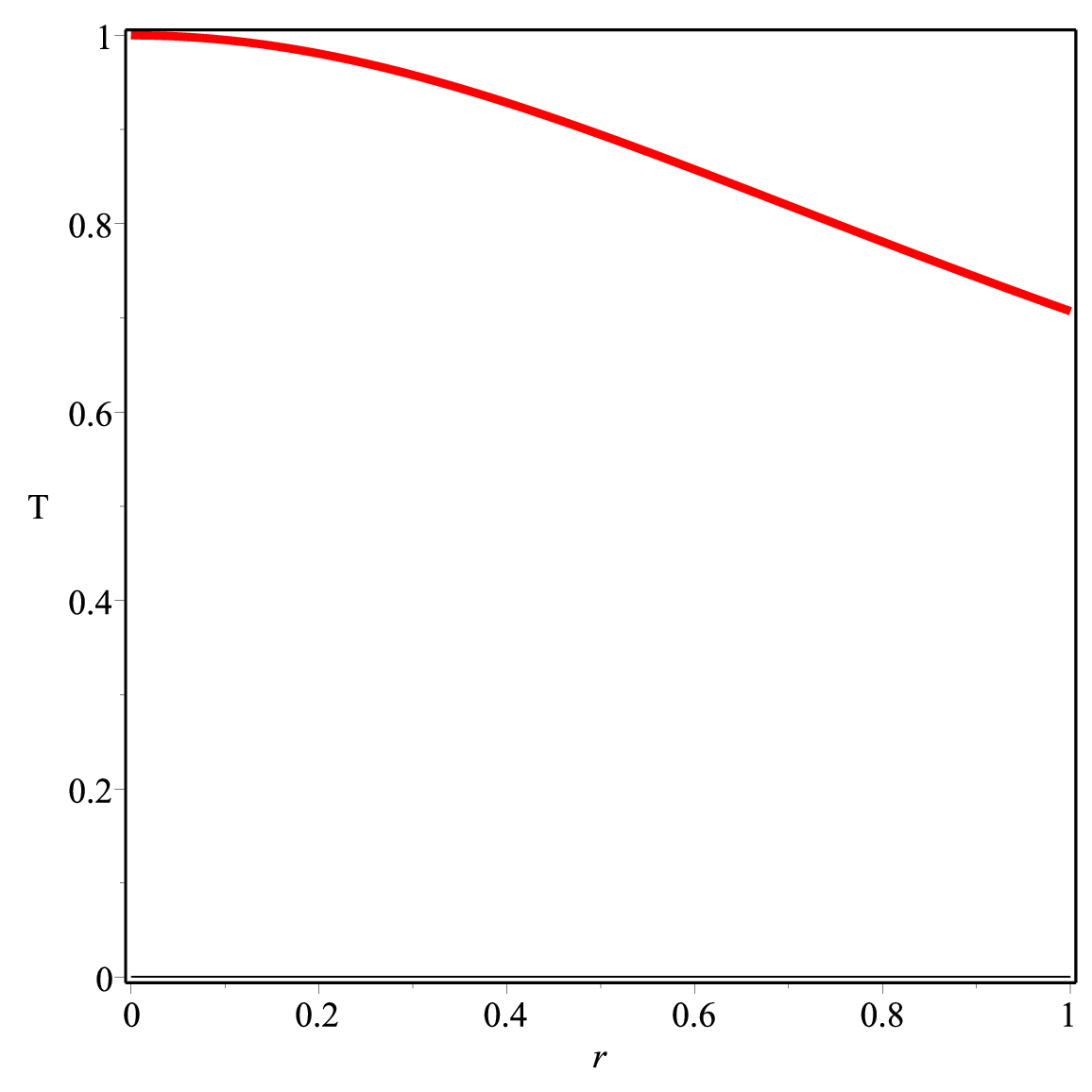}
    \caption{The Tolman temperature as a function of the radial distance $r$.}
    \label{fig1a}
\end{figure}

To investigate the thermodynamic properties of the configuration, it is necessary to determine the temperature profile inside the stellar interior. In the present analysis, the temperature is obtained from the Tolman equilibrium condition (\ref{temp}), which in this case leads to 
\be
T(r) = \frac{T_\infty} {B\sqrt{1+r^2/A^2}}.
\ee 
The temperature therefore decreases monotonically outward from the stellar center (see Fig. \ref{fig1a}). 
This behavior is a direct consequence of the gravitational redshift and represents the relativistic modification of thermal equilibrium in curved spacetime. 
The Tolman temperature guarantees the absence of heat flow and vanishing entropy production throughout the stellar interior. Consequently, the fluid configuration represents a genuine equilibrium thermodynamic system in curved spacetime.

\begin{figure}
    \centering
    \includegraphics[scale=0.4]{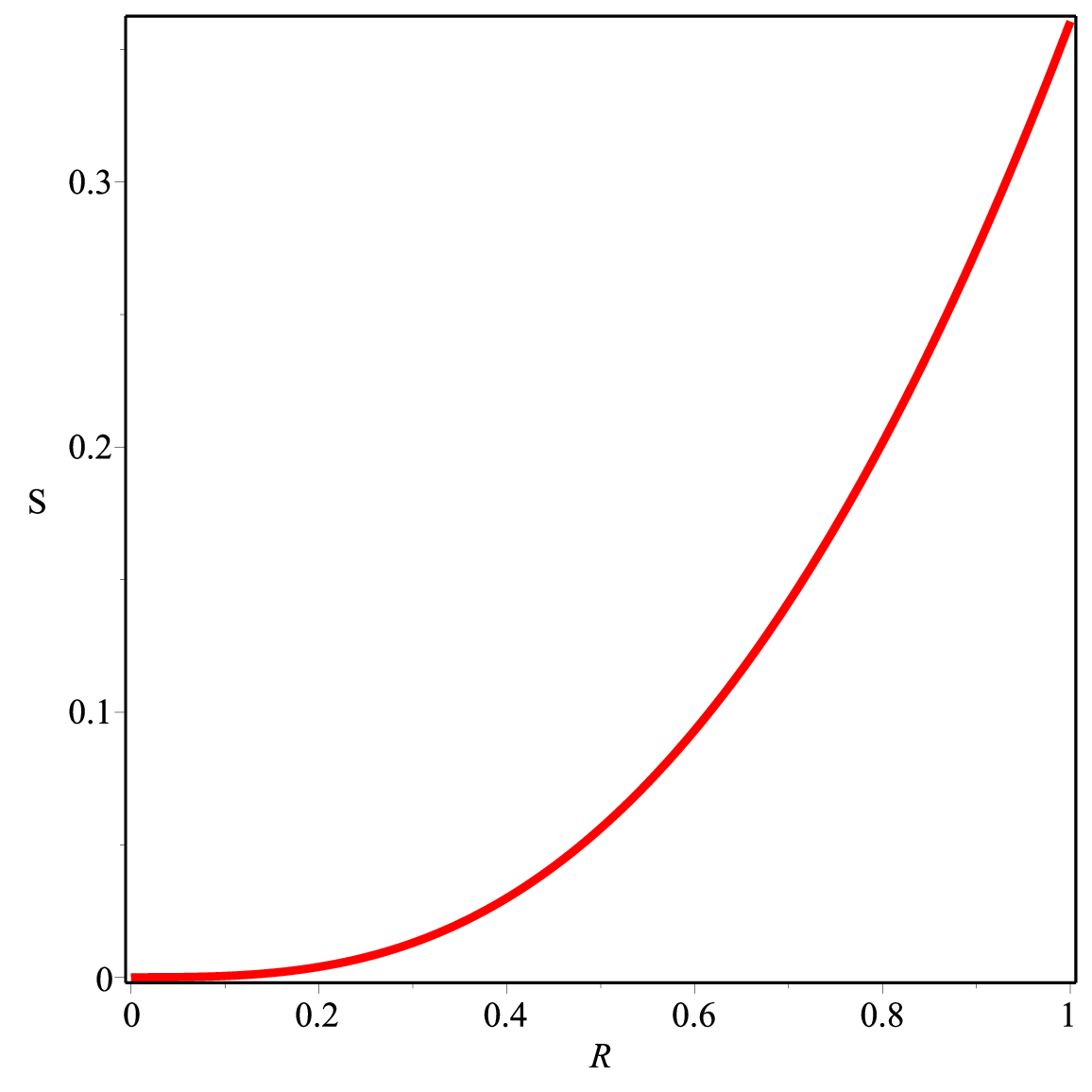}
    \caption{Behavior of the normalized ($A=1,\ B=1,\ T_\infty=1)$ entropy functional in terms of the radius of the object $R$.}
    \label{fig1}
\end{figure}

Once the temperature profile has been specified, the entropy density may be constructed from the relativistic Gibbs relation and the expression (\ref{entropy}).  
Substituting the Tolman IV metric functions and matter variables yields
\be
S= \frac{3B(A^2+2R^2)}{AT_\infty \sqrt{A^2+3R^2}}\int_0^R \frac{(A^2 +r^2) \, r^2}{\sqrt{A^2+3R^2-r^2} \, (A^2+2r^2)^{3/2}} dr.
\ee
The integrand remains finite throughout the stellar interior, and the entropy density is positive whenever the pressure and density remain positive. Consequently, the total entropy is finite and regular for physically acceptable parameter choices. To illustrate the behavior of the entropy, we consider a normalized configuration defined by $A=B=T_\infty=1. $ 
Numerical evaluation of the resulting integral shows that the entropy remains finite and positive throughout the stellar interior, as illustrated in Fig. \ref{fig1}.
In addition, the entropy increases monotonically with the stellar radius and scales approximately as $S\sim R^3$
for sufficiently large radii, consistent with the extensive behavior expected for ordinary thermodynamic systems.

\subsection{Equation of state and thermodynamic interpretation}

The Tolman IV configuration admits an effective equation of state because both pressure and density are monotonic functions of the radial coordinate. Formally eliminating the radial coordinate yields a relation of the form $p=p(\rho)$. 
However, this equation of state is not derived from an underlying microscopic particle model. Instead, it emerges geometrically from Einstein’s equations together with the chosen metric ansatz. Consequently, the resulting fluid does not generally coincide exactly with familiar forms of matter such as radiation fluids, ideal gases, van der Waals systems, or realistic neutron-star equations of state.

Nevertheless, the configuration may still be interpreted as an effective relativistic fluid in equilibrium. The thermodynamic interpretation depends crucially on the choice of temperature profile. In the present work, the Tolman temperature is employed because it guarantees equilibrium in curved spacetime.

As mentioned in previous section, we could alternatively define temperature through a local microphysical relation such as the ideal-gas law,
\be
T = \frac{p}{nk_B}.
\ee 
To determine the explicit integrand of the entropy functional, we assume that $n(r)=n_0\rho(r), \ n_0=const$, which 
may be interpreted as a simple phenomenological relation between the particle number density and the energy density of the fluid. Then, the temperature function becomes

\be
T= {\frac { \left( {R}^{2}-{r}^{2} \right)  \left( {A}^{2}+2\,{r}^{2}
 \right) }{{\it n_0}\,{\it k_B}\, \left( 2\,{A}^{4}+(R^2+r^2)(3A^2+2r^2) \right) }}.
\ee

\begin{figure}
    \centering
    \includegraphics[scale=0.4]{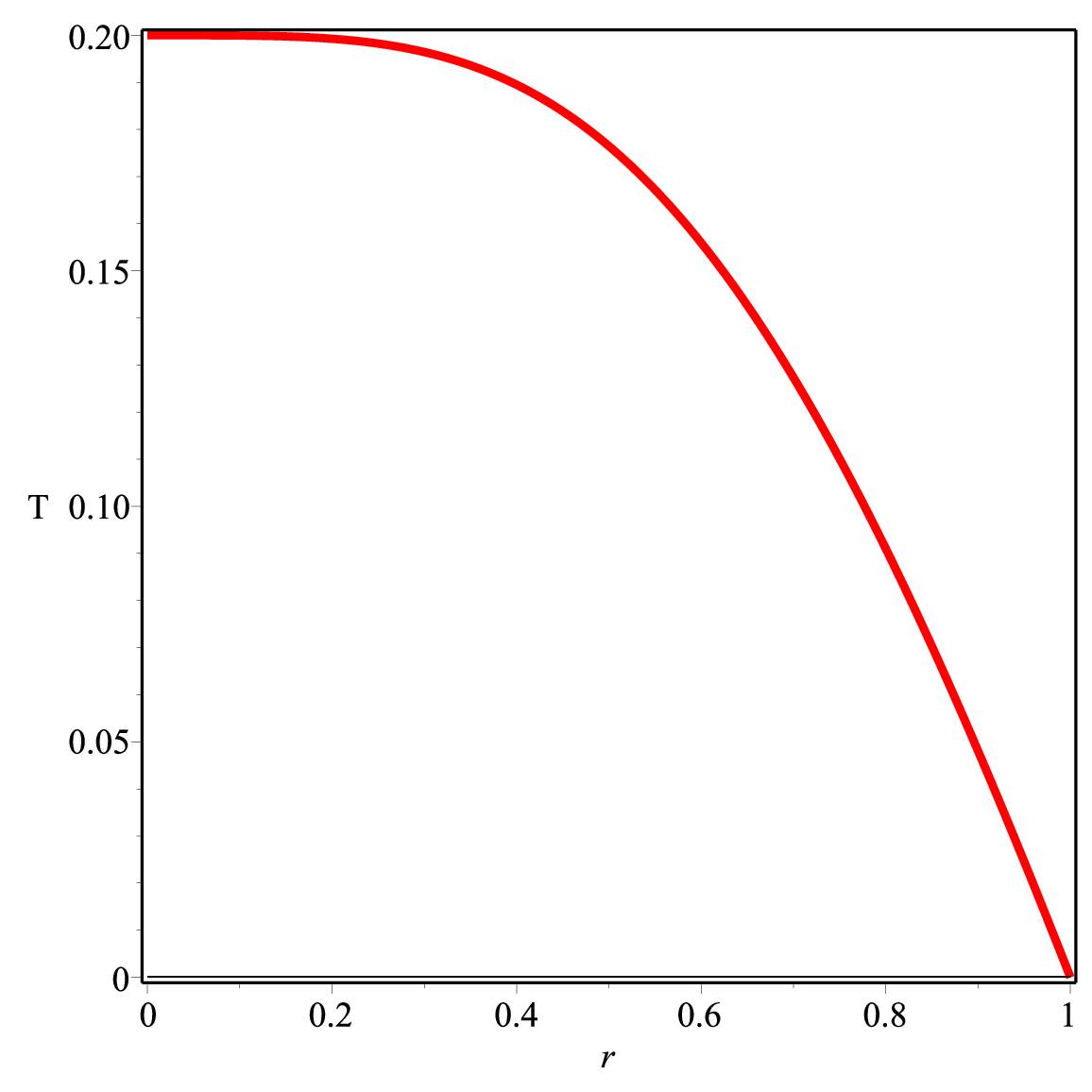}
    \caption{The microscopic ideal-gas temperature in terms of the radial distance $r$. Here we set $n_0=1$, $k_B=1$, $A=1$, and $R=1$.}
    \label{fig2a}
\end{figure}

In Fig. \ref{fig2a}, we depict the behavior of this temperature function.
Although such a temperature may remain locally meaningful, the resulting profile generally does not satisfy the Tolman equilibrium condition. 
In particular, notice that the temperature vanishes at $r=R$ as a consequence of the surface condition for the pressure $p(R)=0$. 
In that situation, the configuration no longer corresponds to a system in exact thermal equilibrium. Heat transport may occur, entropy production becomes nonvanishing, and the associated entropy functional must therefore be interpreted as a non-equilibrium entropy rather than an equilibrium one. 

The entropy functional can be derived from the above expression for the ideal-gas temperature and we obtain 

\be
S= \frac{3n_0k_B(A^2+2R^2)}{\sqrt{A^2+3R^2}}\int_0^R
{\frac { \left( 2\,{A}^{4}+(R^2+r^2)(3A^2+2r^2) \right) \sqrt{A^2+r^2}}{(R^2-r^2) \left( {A}^{2}+2\,{r}^{2}
 \right) ^{5/2}\sqrt {{A}^{2}+3\,{R}^{2}-{r}^{2}}}}
r^2 dr .
\ee

\begin{figure}
    \centering
    \includegraphics[scale=0.4]{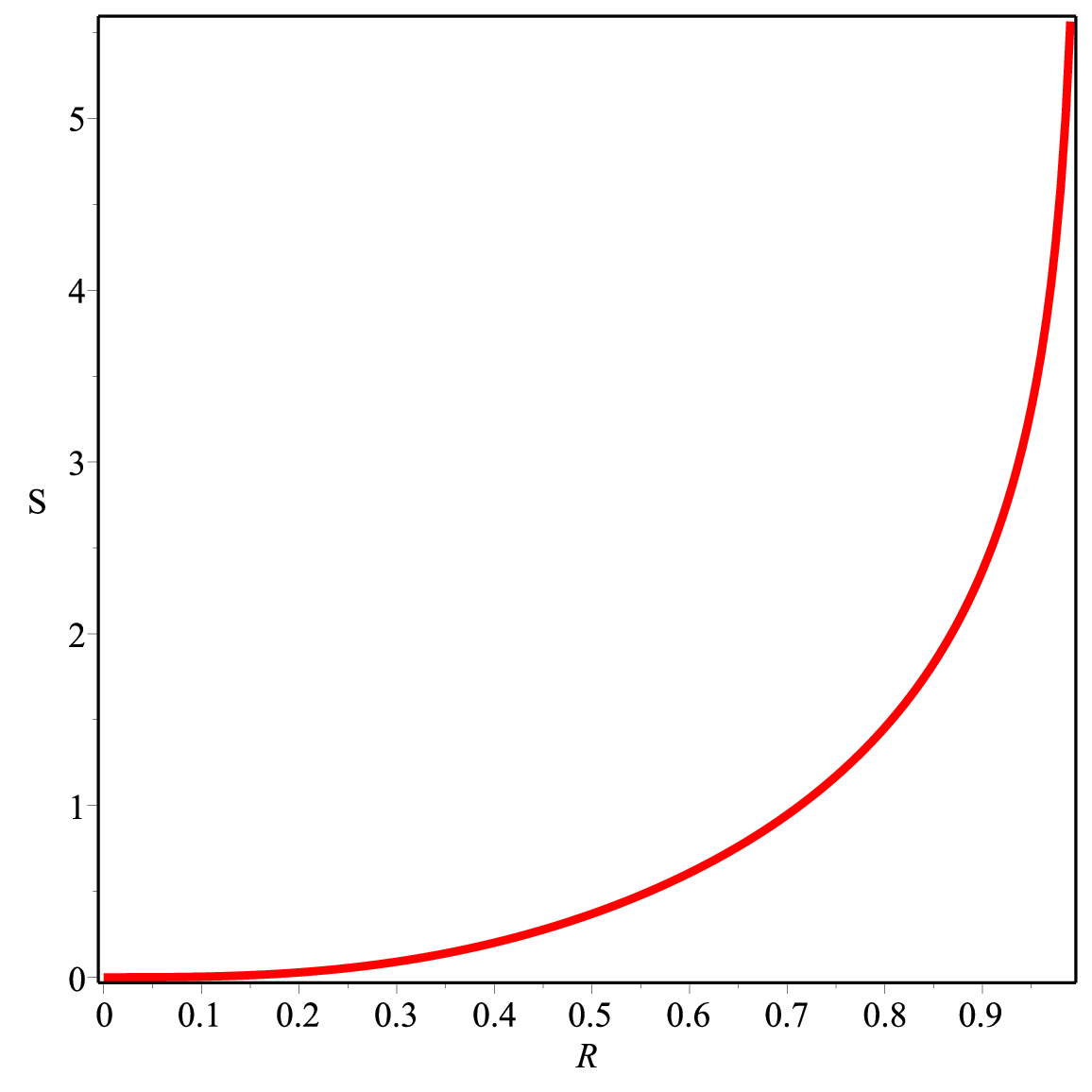}
    \caption{The entropy profile in terms of the radius $R$. Here, we set $A=1,\ n_0=1 $, $k_B=1$, and $R=1$ for concreteness. The integration does not include the surface radius $r=R$ to avoid the singularity in the entropy. }
    \label{fig2}
\end{figure}

In Fig. \ref{fig2}, we illustrate the behavior of the entropy. We see that the entropy profile  remains monotonic increasing with radius, reflecting the cumulative growth of enclosed matter. However, unlike the equilibrium Tolman entropy, it does not remain finite at the boundary. Instead, the integrand develops a non-integrable divergence at $r=R$, leading to $S\to \infty$ as $r\to R$. 
This divergence is not a geometric pathology of the Tolman IV spacetime itself, but rather a consequence of using a non-equilibrium temperature definition that is incompatible with the vanishing-pressure boundary condition. In this sense, the ideal-gas entropy reflects a thermodynamic inconsistency at the stellar surface, where the assumptions underlying local equilibrium break down when extended globally without the Tolman redshift correction.

The Tolman temperature therefore possesses a fundamentally different physical interpretation from temperatures derived solely from local equations of state. Its importance lies in the fact that it incorporates the gravitational redshift effects required by relativistic thermal equilibrium.

\subsection{Thermodynamic acceptability of the Tolman IV solution}

The Tolman IV configuration satisfies the thermodynamic acceptability conditions proposed in the present work in several important respects. The entropy density remains positive throughout the stellar interior, while the total entropy is finite and regular. The temperature profile obtained from the Tolman relation is everywhere positive and monotonic, and the equilibrium condition $
T(r)e^{\nu(r)/2}=T_\infty$ 
is satisfied identically.

Because the Tolman temperature guarantees equilibrium in curved spacetime, the configuration possesses vanishing entropy production and no internal heat flow. Furthermore, the thermodynamic variables remain regular throughout the interior region, and no entropy singularities appear at the stellar center or boundary surface.

The principal limitation of the Tolman IV model is therefore not thermodynamic inconsistency itself, but rather the absence of a realistic microscopic equation of state capable of describing actual nuclear matter inside compact astrophysical objects. Nevertheless, the solution provides a valuable example of a relativistic stellar model that simultaneously satisfies many geometric, hydrodynamic, and thermodynamic acceptability conditions.

\section{Conclusions}
\label{sec:con}

In this work, we have investigated the thermodynamic aspects of static, spherically symmetric perfect-fluid solutions of Einstein’s equations and proposed an extension of the classical concept of physical acceptability to include explicit thermodynamic criteria. The motivation for this analysis arises from the fact that many exact interior solutions, although mathematically consistent and sometimes hydrodynamically viable, do not automatically possess a consistent thermodynamic interpretation. Consequently, the physical relevance of relativistic stellar models should not be assessed solely through geometric regularity and hydrostatic stability, but also through their compatibility with the principles of relativistic thermodynamics.

The analysis was based on the framework of equilibrium thermodynamics in curved spacetime. Starting from the relativistic Gibbs relation and the Euler identity for perfect fluids, we constructed the entropy functional associated with static relativistic stellar configurations. Particular emphasis was placed on the role of the Tolman equilibrium condition, which determines the behavior of temperature in a gravitational field and guarantees the absence of heat flow and entropy production in equilibrium configurations.

A central point of the present work is the distinction between local thermodynamic consistency and genuine global equilibrium. Local equations of state, such as the ideal-gas relation, may remain meaningful within infinitesimal fluid elements even when the overall temperature profile fails to satisfy the Tolman relation. In such situations, the resulting system should be interpreted as a non-equilibrium configuration rather than a true equilibrium relativistic fluid. By contrast, the Tolman temperature incorporates the gravitational redshift effects required by equilibrium thermodynamics in curved spacetime and therefore provides a natural basis for defining equilibrium entropy in self-gravitating systems.

Motivated by the physical acceptability program developed by Delgaty and Lake, we formulated a corresponding set of thermodynamic acceptability conditions for relativistic stellar models. These conditions include positivity and regularity of the entropy density, finiteness of total entropy, positivity of temperature, compatibility with the Gibbs relation, absence of entropy singularities, causal sound propagation, monotonic thermodynamic behavior, and satisfaction of the Tolman equilibrium condition. Together with the usual geometric and hydrodynamic requirements, these conditions provide a broader framework for evaluating the physical plausibility of exact interior solutions.

As an explicit example, we analyzed the Tolman IV solution. This configuration is known to satisfy many of the classical physical acceptability conditions and therefore provides a suitable model for investigating thermodynamic consistency. Using the Tolman temperature profile, we derived the associated entropy functional and showed that the entropy density remains positive and regular throughout the stellar interior. The total entropy was found to be finite, monotonic, and well behaved for physically reasonable parameter choices. Furthermore, the temperature profile satisfies the Tolman equilibrium condition identically, implying the absence of heat transport and entropy production.

The analysis demonstrates that the Tolman IV solution constitutes an example of a relativistic stellar model that is not only geometrically regular and hydrodynamically acceptable, but also thermodynamically consistent within the framework of equilibrium general relativity. The principal limitation of the model lies not in its thermodynamic behavior, but rather in the fact that the corresponding equation of state is geometry-induced rather than derived from realistic microphysics. Consequently, the fluid should be interpreted as an effective relativistic matter distribution rather than a fully realistic description of dense nuclear matter inside compact stars.

The results obtained in this work suggest that thermodynamic consistency should be regarded as an essential complement to the traditional criteria of physical acceptability in relativistic astrophysics. In particular, equilibrium temperature behavior and entropy regularity provide additional restrictions capable of distinguishing physically meaningful solutions from purely mathematical ones.

Several directions for future investigation naturally emerge from the present analysis. One important extension would be the inclusion of nonvanishing chemical potential and particle conservation in the thermodynamic description of relativistic fluids. Another possibility is the study of dissipative systems with heat transport, viscosity, and entropy production, where equilibrium thermodynamics must be replaced by relativistic non-equilibrium thermodynamics. It would also be interesting to generalize the present framework to anisotropic fluids, charged compact objects, and numerical solutions of the Tolman--Oppenheimer--Volkoff equations constructed from realistic nuclear equations of state.

More generally, the present work indicates that the study of exact solutions in general relativity benefits from a unified treatment in which geometric structure, hydrodynamic behavior, and thermodynamic consistency are analyzed simultaneously. Thermodynamic acceptability therefore appears to provide a natural and physically significant extension of the classical program of relativistic stellar modeling.

\section*{Acknowledgments}
The authors acknowledge the support by the Ministry of Science and Higher Education of the Republic of Kazakhstan, grant IRN AP23490322.
The work of HQ was supported by UNAM-DGAPA-PAPIIT, grant No. 108225, and Conahcyt, grant No. CBF-2025-I-243.

\end{document}